\documentclass[aps,twocolumn,prl,showpacs]{revtex4}
\usepackage{amssymb}
\usepackage{epsfig}
\usepackage{graphicx}
\usepackage{amssymb}
\usepackage{wasysym}

\begin{document}

\title{Subdiffusive axial transport of granular materials in a long drum mixer}
\author{Zeina S. Khan and Stephen W. Morris}
\affiliation{Department of Physics, University of Toronto, 60 St. George St., 
 Toronto, Ontario, Canada M5S 1A7 }
\date{\today}


\begin{abstract}
Granular mixtures rapidly segregate radially by size when tumbled in a partially filled horizontal drum. The smaller component moves toward the axis of rotation and forms a buried core, which then splits into axial bands. Models have generally assumed that the axial segregation is opposed by diffusion. Using narrow pulses of the smaller component as initial conditions, we have characterized axial transport in the core. We find that the axial advance of the segregated core is well described by a self-similar concentration profile whose width scales as $t^\alpha$, with $\alpha \sim 0.3 < 1/2$. Thus, the process is subdiffusive rather than diffusive as previously assumed. We find that $\alpha$ is nearly independent of the grain type and drum rotation rate within the smoothly streaming regime. We compare our results to two one-dimensional PDE models which contain self-similarity and subdiffusion; a linear fractional diffusion model and the nonlinear porous medium equation.
\end{abstract}


\maketitle

An interesting property of dry granular materials is their tendency to separate by size under a wide variety of flow conditions \cite{Duranspg,ristow}. Granular segregation is widely found in nature, and plagues industrial processes as well. Probably the best controlled and most widely studied example is segregation along the axis of a partially filled, horizontal ``drum mixer" \cite{Oyama,shinbrot,KHprl,KHpre,KHspg,KCprl,KCpre,ottino_prl,our_EPL,chris,ristownak,nakagawa,shattuck}. After hundreds of drum rotations, an initially mixed binary distribution of different-sized grains sorts itself into almost periodic bands along the axis of the drum. These bands are threaded by a radial core of the smaller grains which develops prior to axial band formation \cite{KHprl,KHpre,ristownak,nakagawa,shattuck,our_EPL}.  The radial core typically forms after just a few drum rotations. Accounting for this rich dynamical behaviour has been the goal of cellular automata models \cite{automata}, molecular-dynamics simulations \cite{rapaport} and several continuum theories \cite{Savage,Ziketal,Levitan,Levinechaos,ATprl,ATVpre,elperin}. The axial bands must somehow be sustained against being mixed away by the random motion of the grains.  Continuum models have generally assumed that the random motions mimic normal diffusion, and therefore that normal Laplacian gradient terms determine the short-wavelength cutoff of the axial band pattern. In this Letter, we experimentally challenge this common assumption.  We find, surprisingly, that the axial transport of the radially segregated core along the drum is much slower than diffusion, {\it i.e.} that it is subdiffusive.  It is nevertheless described by a self-similar profile which scales approximately as $t^{1/3}$.  We also find that the self-diffusion of the larger particles is subdiffusive.  These results have strong implications for models of axial segregation, and possibly for other theories of granular mixing.

Early theoretical models regarded axial band formation as the result of a diffusion process with a negative diffusion coefficient \cite{Savage,Ziketal,Levinechaos}. These models ignore the radially segregated core, and they cannot account for the oscillatory transient that precedes axial band formation in some mixtures \cite{KCprl,KCpre,ottino_prl}. This oscillatory travelling wave state apparently demands that the basic dynamics be at least second order in time. A later model due to Aranson {\it et. al} \cite{ATprl, ATVpre} reproduced both axial segregation as well as the oscillatory transient, while still ignoring the core.  We have recently shown experimentally that this model is also inadequate\cite{our_EPL}, and that the core dynamics are themselves oscillatory.   Another model due to Elperin {\it et. al} \cite{elperin} regards axial segregation as resulting from a radial core instability leading to a spatially periodic thickening of the core.  This model, unfortunately, also cannot account for the travelling wave state. In all cases, these models explain the short wavelength cutoff of the axial band pattern as the result of the supposed axial diffusion of the smaller grains. We show experimentally below that axial transport is not well-described by normal diffusion.  This is true of either the smaller grains in a binary mixture or of the larger grains in a self-mixing process. This falsifies a common, basic assumption of segregation models.

A few studies have investigated the axial transport of grains experimentally \cite{old_diff,ristownak,nakagawa,shattuck}, but none have systematically investigated the effects of varying grain type and drum rotation rate. Here we report experiments which characterize the axial transport of radially segregated grains using several different grain types and drum rotation rates, starting with a narrow pulse initial condition.

%

The drum mixer used in all experiments consisted of a horizontal Pyrex tube, 600 mm long with an inner diameter of 28.5 mm, rotated about its long axis at a constant rotation rate of 0.31 $\textrm{rev}/\textrm{s}$ or 0.62 $\textrm{rev}/\textrm{s}$. The larger grains were either cubic white table salt or transparent glass spheres and had a size range of 300-420 $\mu$m. The smaller grains were either irregularly shaped black hobby sand or bronze spheres, with a size range of 177-212 $\mu$m. The filled volume fraction of the drum was 28 \%. In order to reproducably fill the drum, the grains were loaded into a long U-shaped channel, which was inserted lengthwise into the drum and rotated to deposit its contents. 
\begin{figure}[htb!]
\begin{center}
\epsfig{file = 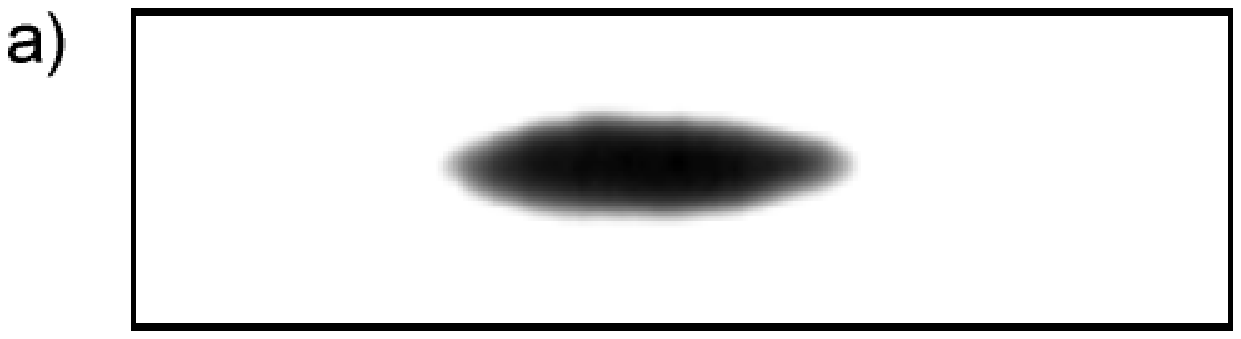,width = 2.25in}
\epsfig{file = 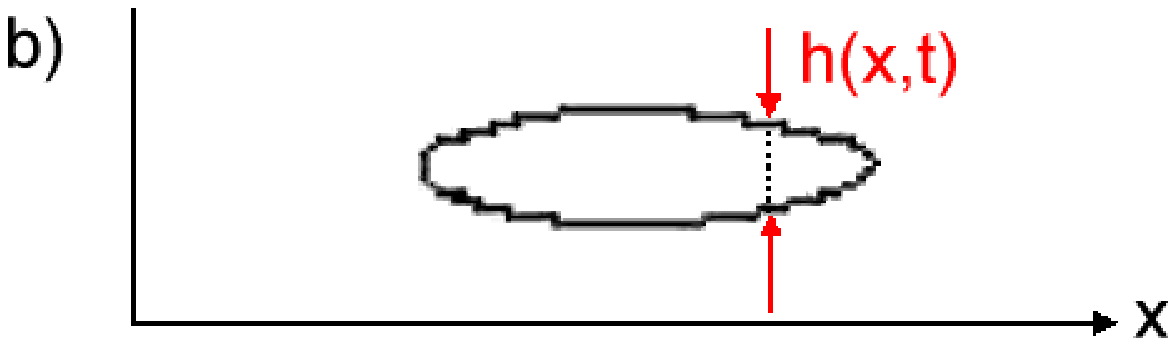,width = 2.25in}
\epsfig{file = 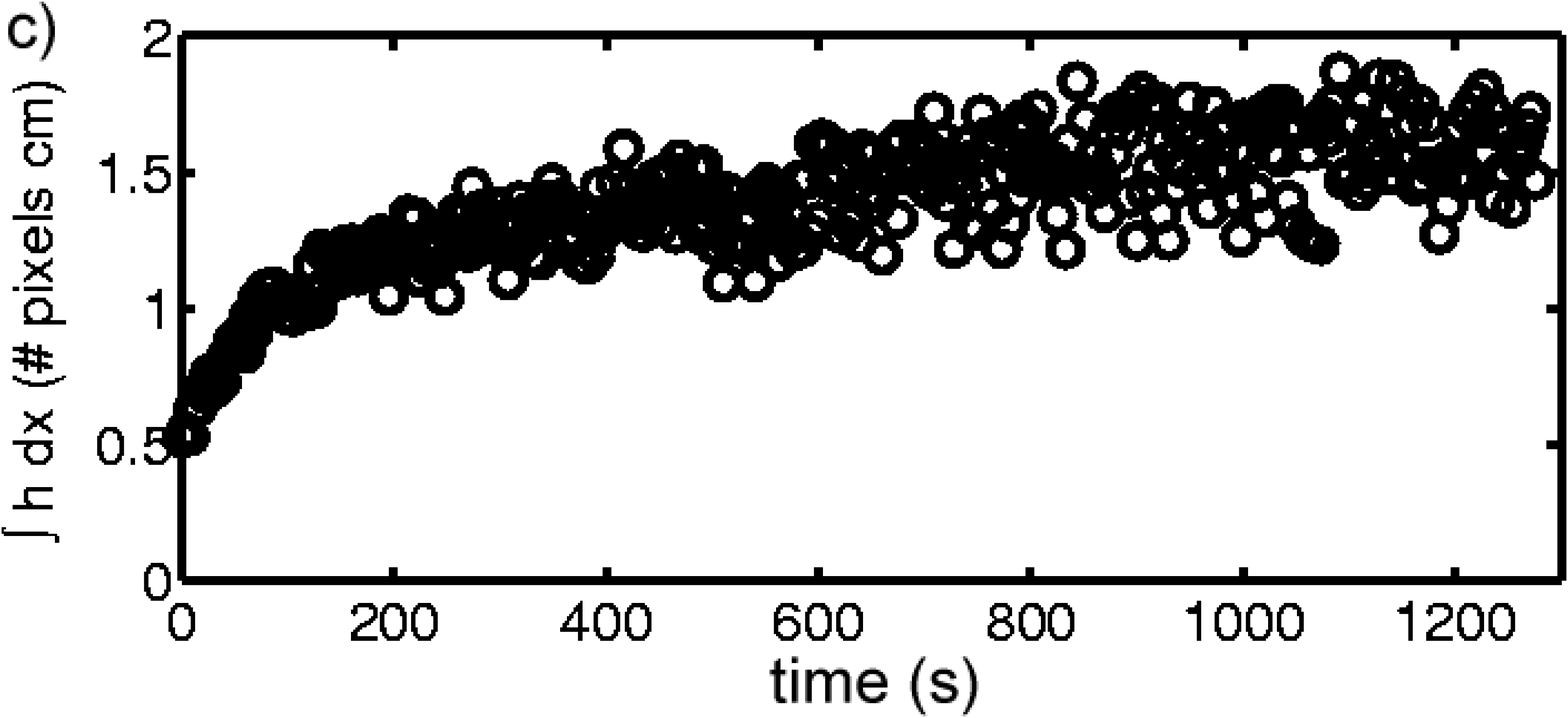,width = 2.25in}
\epsfig{file = 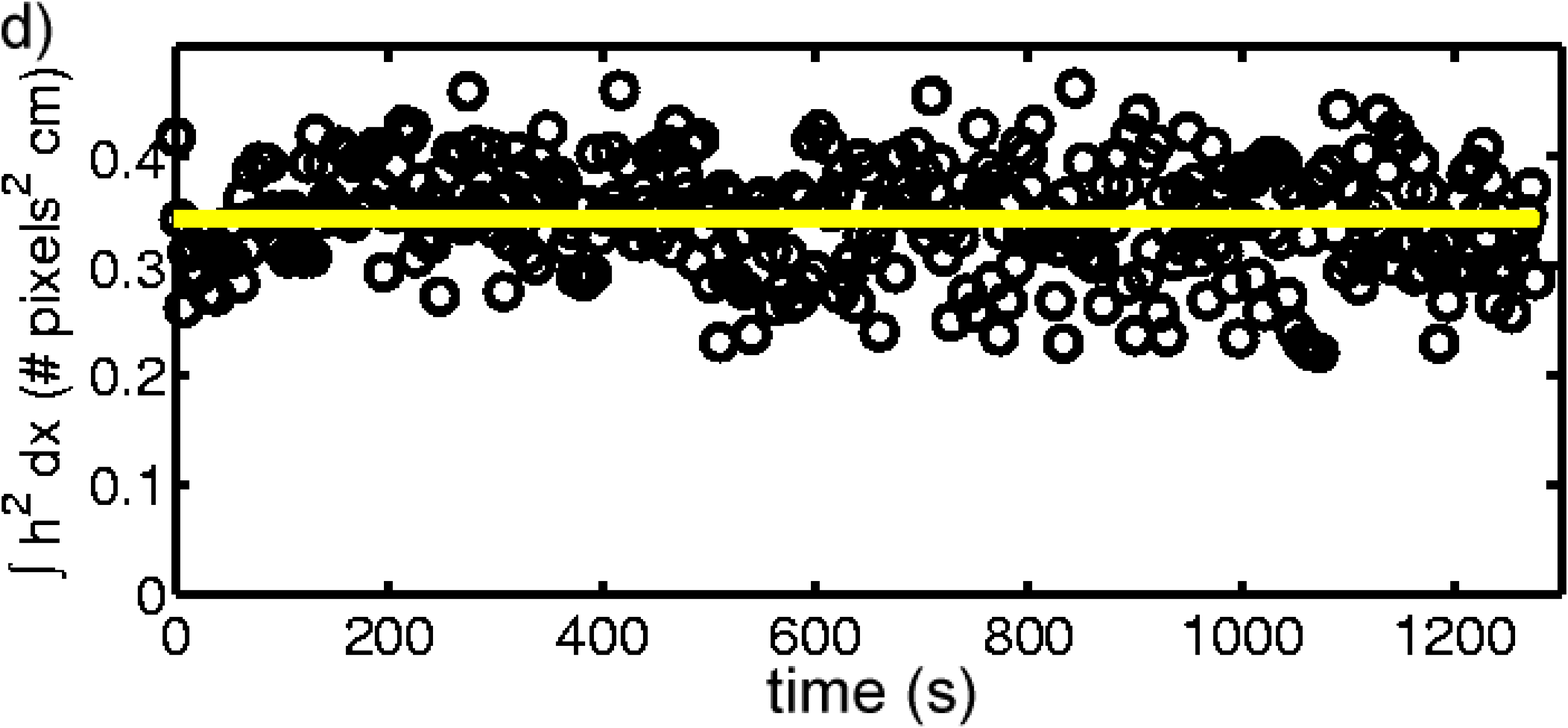,width = 2.25in}
\caption{(Color online) a) An image of the shadow of the radial core formed by 177-212 $\mu$m sand grains  within 300-420 $\mu$m salt grains.  b) The detected edge used to determine the vertical extent $h(x,t)$ of the core.  c) The $x$ integral of $h$ is not constant in time, and is thus not proportional to the volume of small grains contained within the radial core. d) The integral of $h^2$ is constant in time and is proportional to the desired concentration. }
\label{Ffig1}
\end{center}
\end{figure}
\noindent This procedure ensures a uniform filling fraction of the drum. To obtain reproducible, quantitative dynamical information, we used a pulse initial condition. The pulse was made by placing thin spacers in the U-shaped channel 1.5 mm apart. The 1.5 mm space was filled with the smaller grains, and the remaining space was filled with the larger grains.

After a few drum revolutions, the pulse of small grains forms a subsurface radial core and cannot be observed using standard surface lighting and video imaging techniques. Instead, we used a bulk visualisation technique developed by Khan {\it et al}~\cite{our_EPL}. The large grains are transluscent and the small grains are opaque. When a bright light source is placed behind the rotating drum, one can observe a shadow of the radial core on the front face of the granular sample. This shadow is a two-dimensional projection of the radial core. A computer controlled high speed camera was used to observe the radial core shadow. Five images per drum revolution were obtained and averaged to determine the evolution of the radial core. Figure \ref{Ffig1}a shows a typical image of the radial core shadow. Using edge detection, the radial core

\begin{figure}[htb!]
\begin{center}
\epsfig{file = 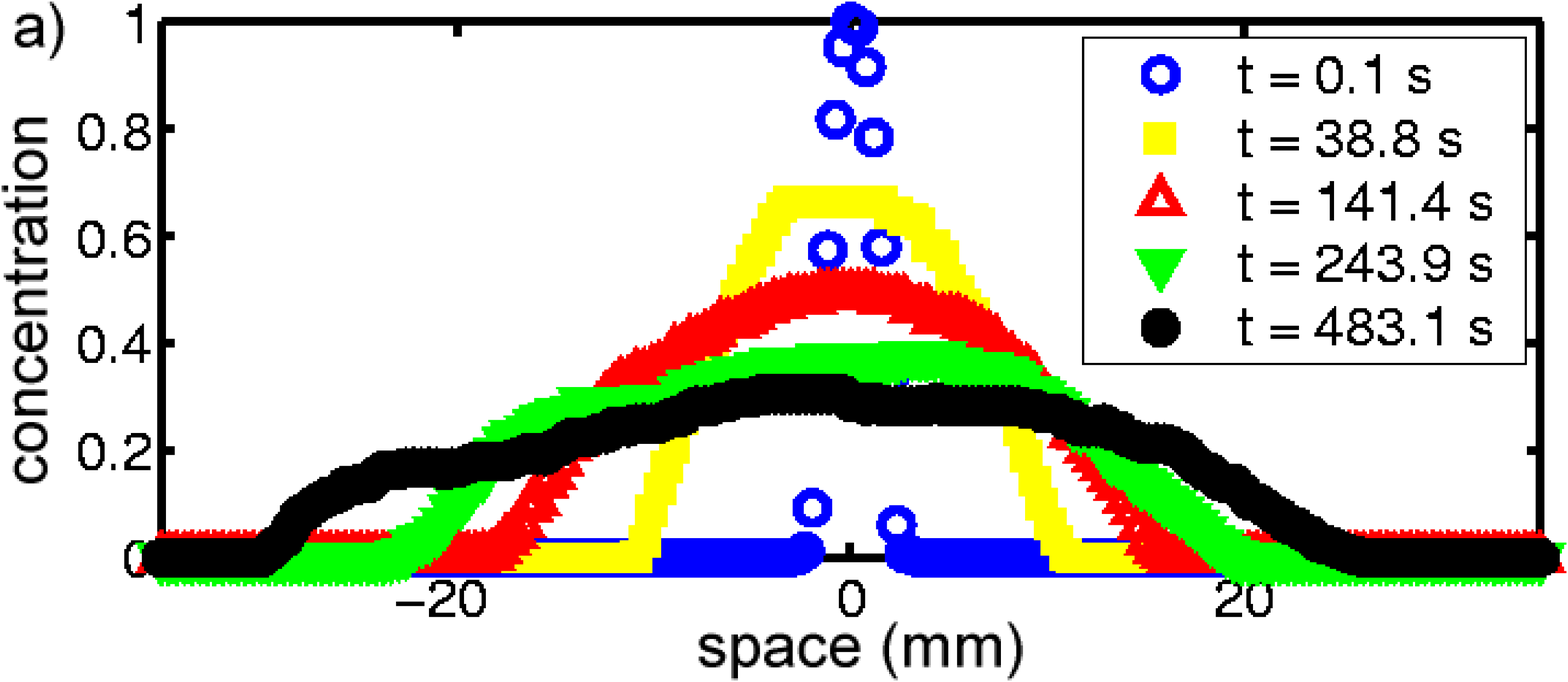,width = 2.5in}
\epsfig{file = 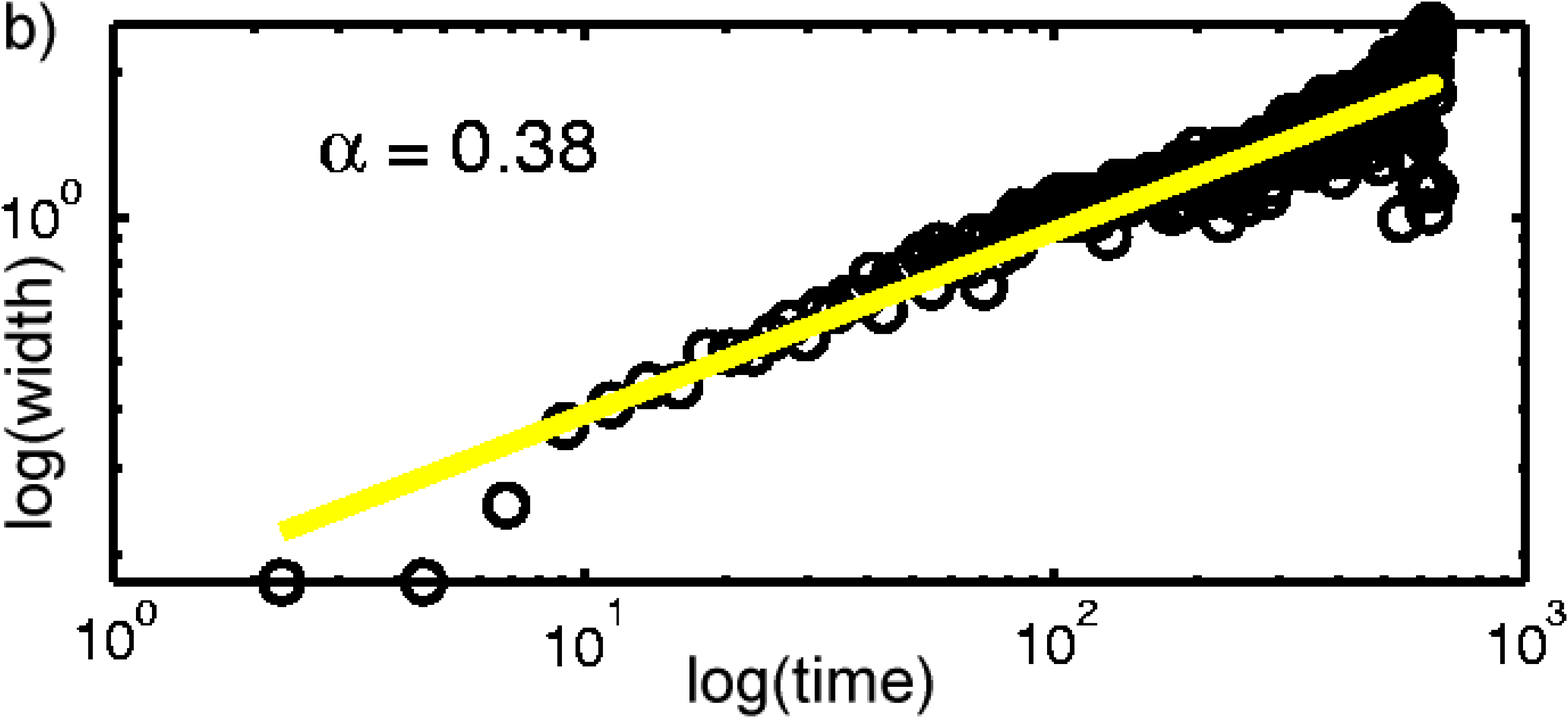,width = 2.5in}
\epsfig{file = 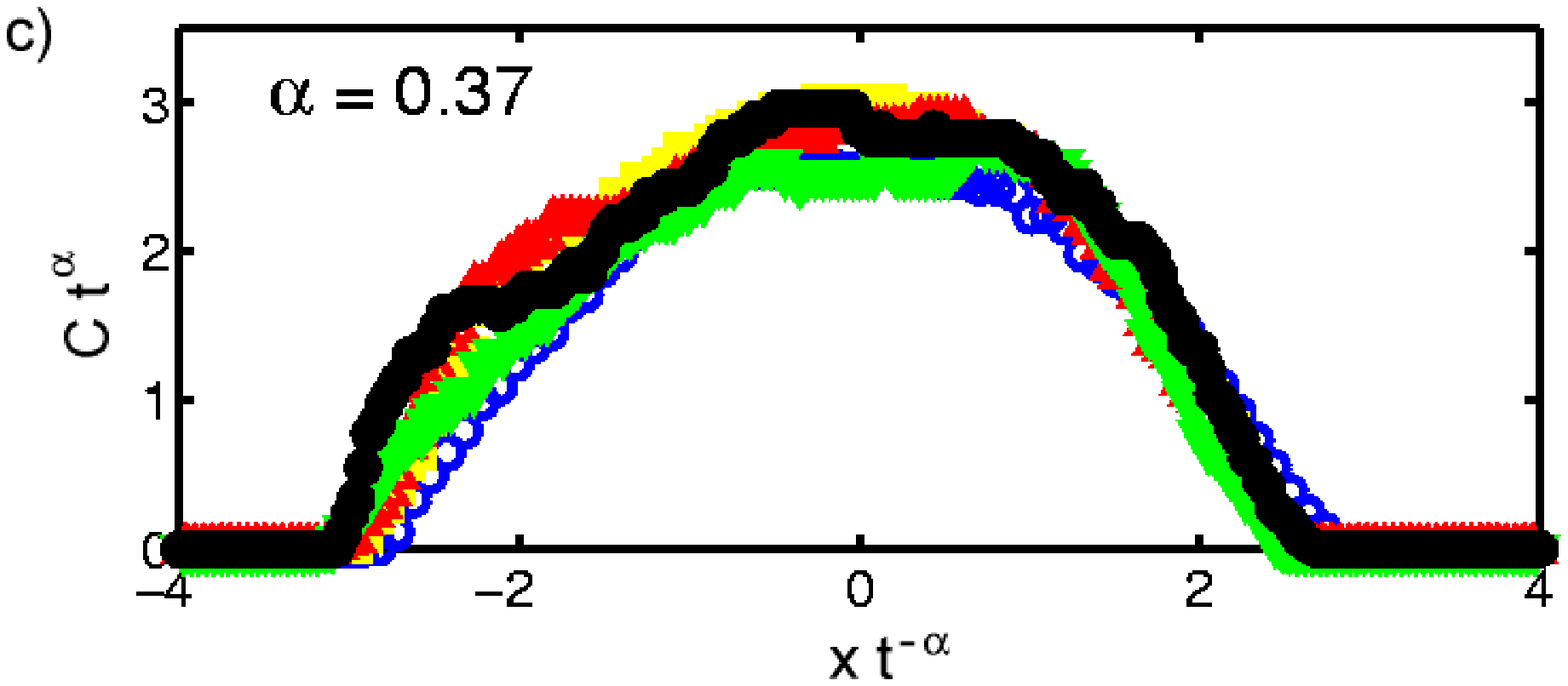,width = 2.5in}
\caption{(Color online) a) Concentration profiles of a spreading radial core pulse of sand grains within salt grains at various times. b) Power-law scaling of the FWHM of the radial core pulse. From the linear fit (yellow line) we find that the width $\propto$ $t^{0.38}$.  c) Collapsed concentration profiles of the radial core pulse corresponding to (a). The collapse parameter is $\alpha = 0.37$. }
\label{Ffig2}
\end{center}
\end{figure}

\noindent height $h(x,t)$ was measured as shown in figure \ref{Ffig1}b, and expressed as a fraction of the full height of the material in the drum. If we assume that any cross section of the three dimensional structure of the radial core is an ellipsoid, the square of the diameter of the radial core $h^2$ is proportional to the volume of small grains contained in the radial core at each axial position $x$. Figure \ref{Ffig1}c shows

\begin{table}
\caption{\label{Ttable1}Collapse parameters for the self-similar spreading of radial cores in various grain types and rotation frequencies.}
\begin{tabular}{|c |c |c |c|}\hline Large grains&Small grains&Rotation rate& $\alpha$\\ 300-420 $\mu m$ & 177-212 $\mu m$ & $ (\textrm{rev}/\textrm{s})$ &
\\\hline salt	& sand & 0.31 &	0.38 $\pm$ 0.03\\\hline salt	& sand & 0.62 &	0.37 $\pm$ 0.03
\\\hline glass & bronze &	0.31 & 0.31 $\pm$ 0.04\\\hline glass & bronze &	0.62 & 0.29 $\pm$ 0.01
\\\hline glass & sand & 0.31 & 0.35 $\pm$ 0.03\\\hline glass & $-$ & 0.31 & 0.34 $\pm$ 0.04
\\\hline salt & $-$ & 0.31 & 0.29 $\pm$ 0.01\\\hline
\end{tabular}
\end{table}

\vspace{0.3in}

\noindent   the time evolution of the $x$ integral of $h$, which increases with time. Figure \ref{Ffig1}d shows that the $x$ integral of $h^2$ is constant in time, as it should be for a conserved quantity.   This validates our assumption about the core shape, and demonstrates that $h^2$ can be used as a local concentration measure. The error in measurement of $h^2$  corresponding to an error in $h$ of $\pm$ 2 pixels.

%

For a normal diffusive process, the width of a narrow pulse initial condition grows as $t^{1/2}$. In our experiment, the pulse of small grains do not mix into the larger ones, but instead the pulse sinks below the surface of the larger grains forming a radial core, which then spreads axially.  We can nevertheless ask if this axial spreading is analogous to normal diffusion, as is assumed in  models\cite{Savage,Ziketal,Levinechaos,ATprl,ATVpre,elperin}.  Figure \ref{Ffig2}a shows the radial core concentration profile at different times, for a mixture of   small sand grains and large salt grains. Plotting the full-width at half-maximum of the concentration profile against time, we determined the power-law dependence of the radial core width with time, as shown in figure \ref{Ffig2}b.  From this, we determined that the width scales as $t^{\alpha}$, where $\alpha < 1/2$. This analysis, however, only determines the power-law time dependence of one arbitrarily chosen dimension of a pulse (here, the half-maximum width) and not the whole pulse shape. For a symmetric initial condition, data collapse can test the scaling of the entire pulse. Figure \ref{Ffig2}c shows collapsed data corresponding to the concentration profiles in \ref{Ffig2}a, where the axial length scale was transformed as $x \rightarrow {x}{t^{-\alpha}}$ and the axial concentration of small grains $C(x,t)$ was transformed as $C \rightarrow {C}{t^{\alpha}}$. The pulse width increases at the same rate as the pulse amplitude decreases, thus the spreading process is self-similar. This implies that the integrated concentration is constant and that no grains are lost from the core. The average collapse parameter for large salt grains and small sand grains with a drum rotation rate of 0.62 $\textrm{rev}/\textrm{s}$ is $\alpha = 0.37 \pm 0.03$, averaged over 10 runs. Similar experiments were repeated for different combinations of grains at two drum rotation rates.  The results are shown in table 1 \ref{Ttable1}.  We conclude that cores of small grains spread axially as $t^{\alpha}$ where $\alpha \sim 1/3 < 1/2$, independent of grain type and drum rotation rate within the smoothly streaming regime.

\begin{figure}[htb!]
\begin{center}
\epsfig{file = 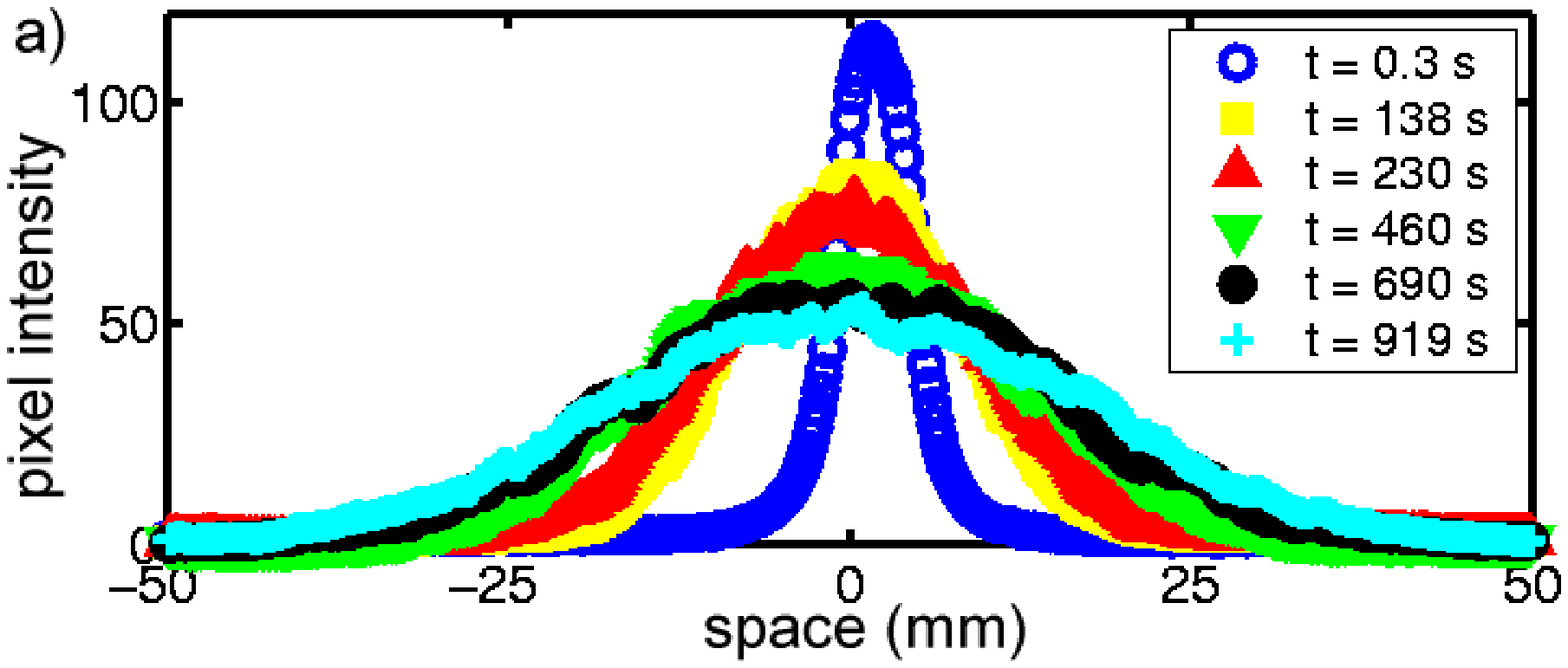,width = 2.5in}
\epsfig{file = 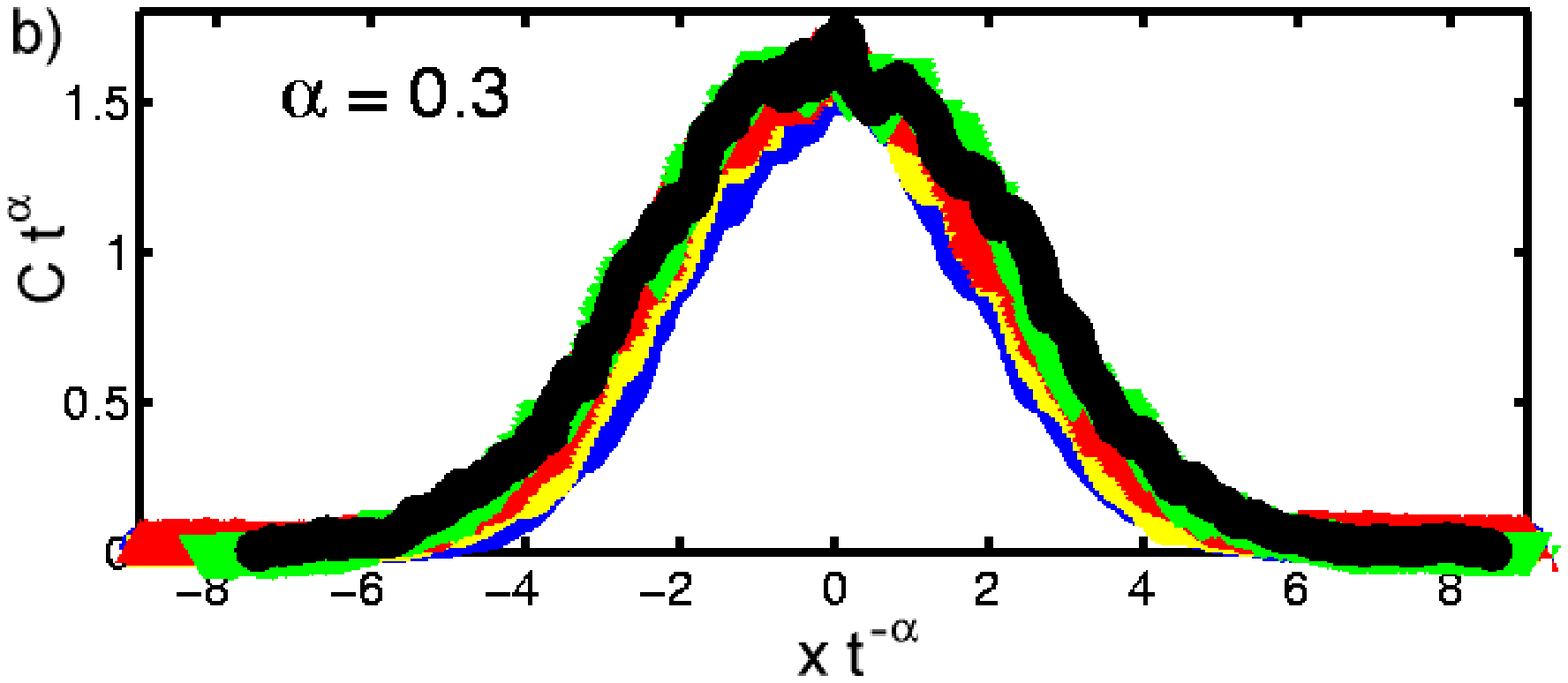,width = 2.5in}
\caption{(Color online) a) Concentration profiles of a mixing pulse of dyed black salt grains surrounded by white salt grains. b) Collapsed concentration profiles corresponding to a). The collapse parameter is $\alpha = 0.3$. }
\label{Ffig3}
\end{center}
\end{figure}

It is interesting to compare the spreading of radially segregated cores of small grains with the non-segregating self-diffusion of the large grains alone.  To observe this experimentally, some of the large grains were dyed black. These dyed grains were loaded into a drum full of otherwise identical white grains with a 1.5 mm wide pulse as the initial condition.  The space-time evolution was observed using standard surface-lighting and imaging techniques \cite{KCprl,KCpre}. Figure \ref{Ffig3}a shows the concentration profile of dyed salt particles at various times. This data was collapsed in a similar way as discussed previously, as shown in figure \ref{Ffig3}b.  Again, we find a collapse parameter $\alpha < 1/2$.  For runs using salt grains, $\alpha = 29 \pm 0.01 $ and for runs using glass spheres, $\alpha = 34 \pm 0.04 $, each averaged over 5 runs.  Thus, we conclude that the self-diffusion of grains in the rotating drum is also subdiffusive, even when no segregation is involved.  We discuss some differences between these two cases below.

\vspace{-0.4in}

In addition to examining the temporal scaling of the pulse, we can also measure in detail the functional shape of the scaling solution.  Here is it possible to distinguish between different subdiffusive processes. We have investigated two candidate models for radial core spreading; the fractional diffusion equation (FDE) and the porous medium equation (PME). The fractional diffusion equation is 
\begin{equation}
\frac{\partial^{\gamma}}{\partial t^{\gamma}} C(x,t) = D\frac{\partial^2}{\partial x^2}C(x,t),
\end{equation}
 where $\gamma = 2\alpha$ denotes the order of a fractional time derivative\cite{wyss,metzler}. Solutions of this linear equation have the property that the width of a narrow pulse initial condition grows as $t^{\alpha}$, where $\alpha \le 1/2$. If $\alpha = 1/2$, the solution reduces to normal Fick diffusion. This FDE model is often used to describe processes which occur in spaces where there are temporal or spatial constraints, such as the flow of tracers through porous media \cite{henry}. The FDE has an analytic series solution in terms of Fox's H-Functions \cite{wyss,metzler}, which forms the self-similar scaling solution. We also examined the porous medium equation (PME),
\begin{equation}
\frac{\partial}{\partial t}C(x,t) =\tilde D\frac{\partial^2}{\partial x^2}(C(x,t)^2).
 \end{equation}
This nonlinear model describes the spreading of a compact mound, and has the property that for a narrow pulse initial condition, the width grows as $t^{1/3}$, and the scaling solution has a parabolic profile\cite{barenblatt}.

\begin{figure}[htb!]
\begin{center}
\epsfig{file = 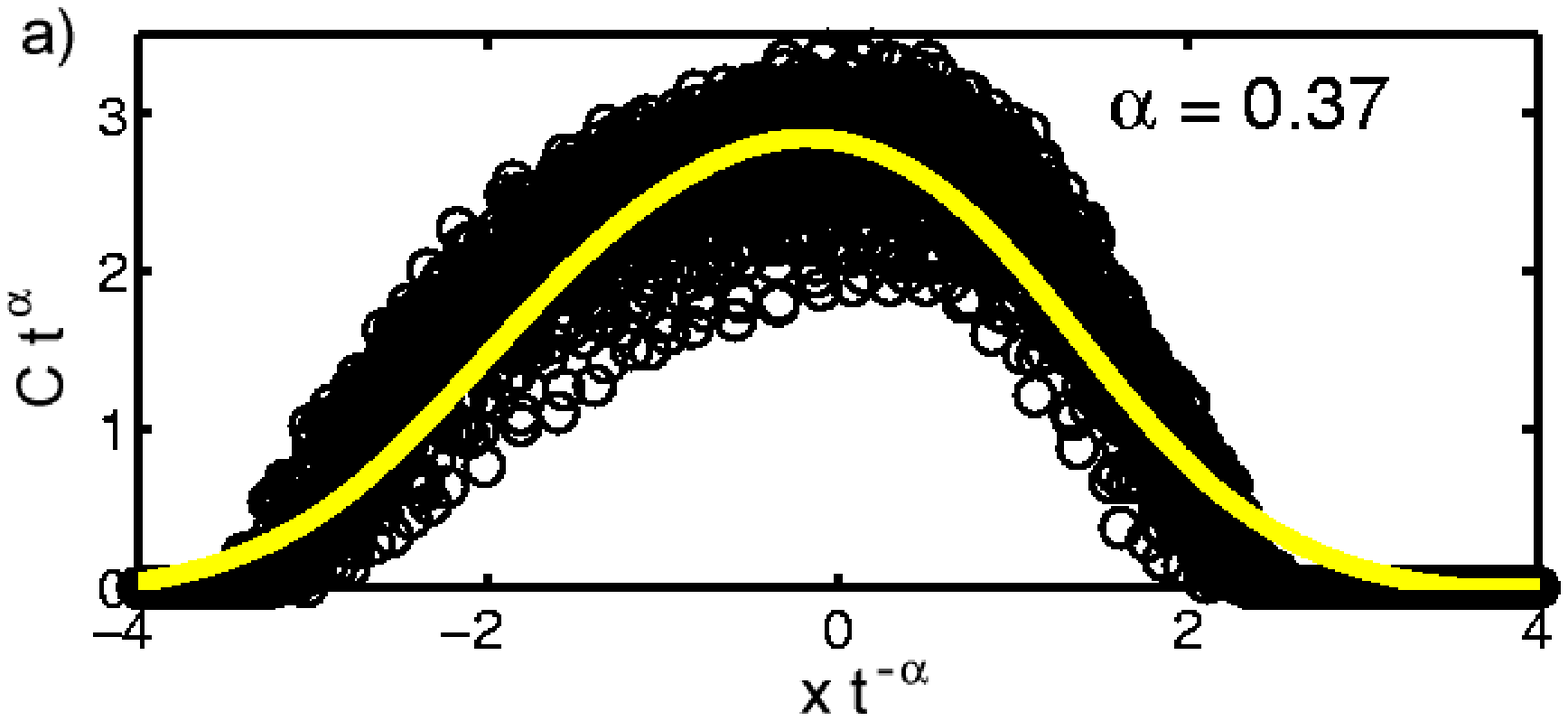,width = 2.25in}
\epsfig{file = 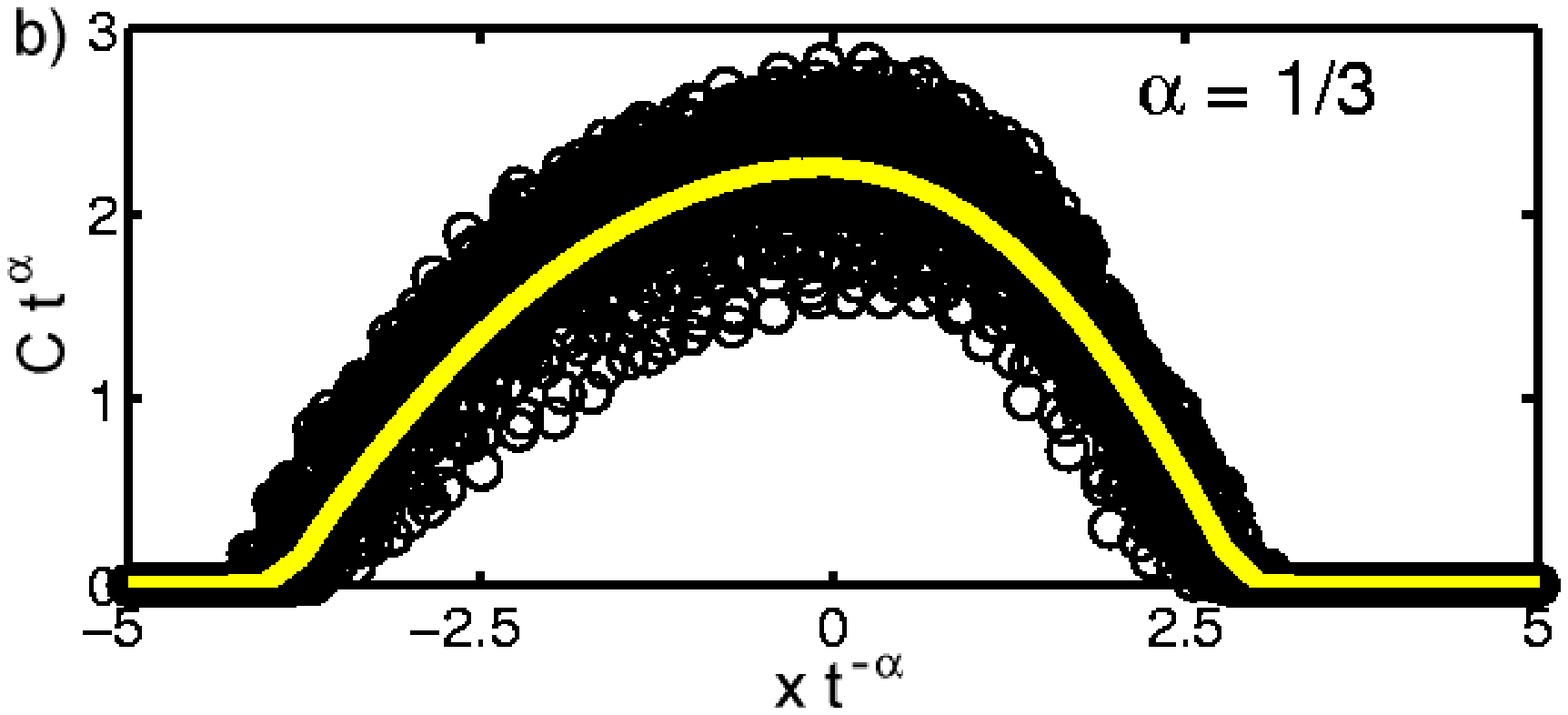,width = 2.25in}
\epsfig{file = 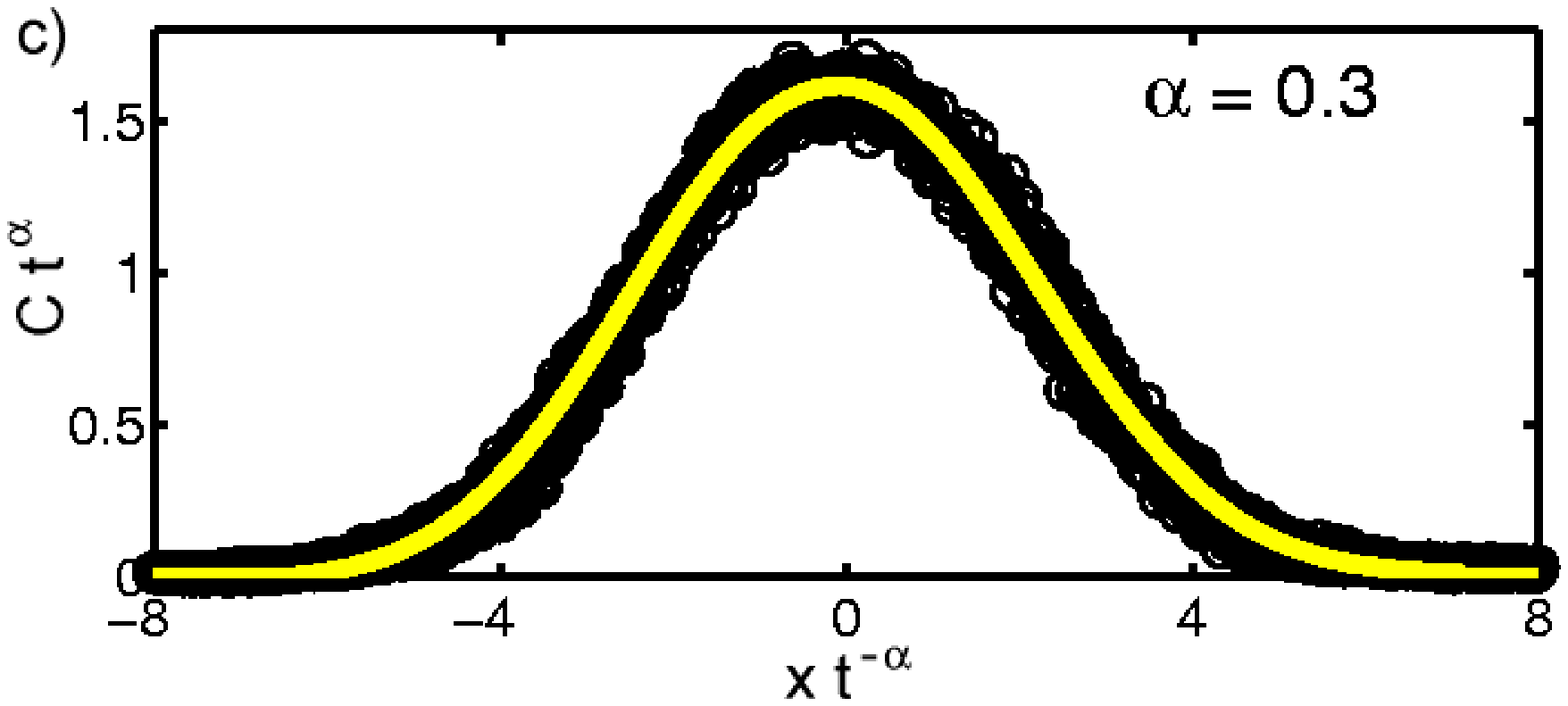,width = 2.25in}
\epsfig{file = 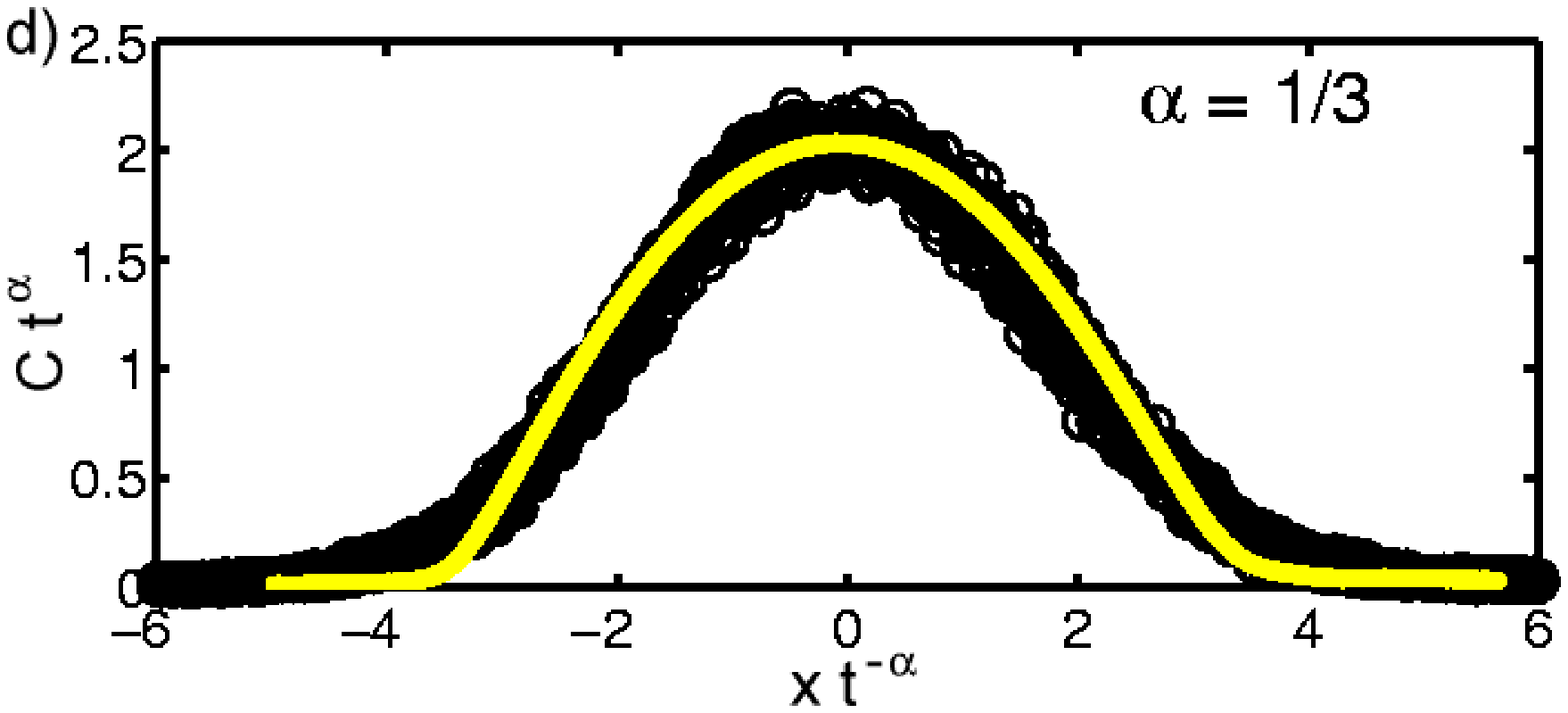,width = 2.25in}
\caption{(Color online) Collapsed concentration profiles of a black sand radial core in salt grains fit to: a) the fractional diffusion equation (yellow line), b) the porous medium equation (yellow line). Collapsed concentration profiles of mixing dyed black salt grains fit to: c) the fractional diffusion equation (yellow line), d) the porous medium equation (yellow line). }
\label{Ffig4}
\end{center}
\end{figure}

 We fit radial core concentration data collapsed with $\alpha$ as a free parameter to the series solution of the FDE, and data collapsed with $\alpha = 1/3$ to a numerical solution of the PME, as shown in figure \ref{Ffig4}a and b respectively. We find that while both solutions model the collapsed concentration profiles reasonably well within experimental error, the PME has a smaller systematic discrepancy, since the profiles are better described as parabolic.  The FDE solution has exponential wings and inflection points that are not obvious in the data.  We note, however, that our projection visualization technique may simply be too insensitive to detect these tails. 
 
 We also fit the non-segregating self-diffusion of the large grains to both models and find that the FDE gives a qualitatively better fit because in this case, the concentration profiles have tails within experimental resolution, while the parabolic PME solution does not. Examples of these fits to collapsed concentration profiles of mixing salt grains are shown in figures \ref{Ffig4}c-d.   In all cases, however, fits to ordinary Fick diffusion with $\alpha = 1/2$ are very poor.

%
 
In conclusion, our results show that the axial transport of grains in a rotating tube is a subdiffusive process.  This is true of both small particles comprising a segregated radial core as well as for  surface mixing of larger grains.  In all cases, we find temporally self-similar concentration profiles that scale approximately as $t^{1/3}$.  These conclusions suggest that spontaneous axial segregation patterns in such tubes are more weakly damped, in the sense that they are sustained against slower mixing processes, than has been previously supposed.    The goal of our future work is to elucidate the connection between axial band formation and the axial transport of grains, which is still unclear. 

\acknowledgments
We wish to thank Wayne Tokaruk, Mary Pugh and Frank Van Bussel. This work was supported by the Natural Science and Engineering Research Council of Canada.

\end{document}